\documentclass[journal]{IEEEtran}

\usepackage{amsmath,amssymb,amsfonts}
\usepackage{algorithmic}
\usepackage{graphicx}
\usepackage{textcomp}
\usepackage{xcolor}
\usepackage{booktabs}
\usepackage{multirow}
\usepackage{url}

\begin{document}

\title{An AI Security Agent for University ACMIS: Multi-Vector
Threat Detection and Automated Response}

\author{Joseph Walusimbi~and~Joshua Benjamin Ssentongo

\thanks{J. Walusimbi and J. B. Ssentongo are with the
Department of Electronics and Computer Engineering,
Soroti University, Soroti, Uganda.
(e-mail: 2401600068@sun.ac.ug; j.ssentongo@sun.ac.ug).}

\thanks{Corresponding author: J. Walusimbi
(e-mail: 2401600068@sun.ac.ug).}}

\maketitle

\begin{abstract}
University Academic Management Information Systems (ACMIS) are
high-value targets for a wide spectrum of security threats including
brute-force login attacks, payment fraud, privilege escalation, insider
data theft, and academic integrity violations. Traditional rule-based
intrusion detection systems are inadequate because many malicious
activities are structurally indistinguishable from normal operations.
This paper presents an AI-based security agent for ACMIS that combines
supervised anomaly detection, behavioural analytics, and a natural
language processing chatbot for secure password recovery. The agent
monitors five operational layers---authentication, authorisation,
financial transactions, user behaviour, and system health---and responds
through a four-tier risk escalation framework. A modular architecture
allows the core engine to be extended to other institutional systems.
Experiments on a simulated ACMIS event log dataset of 147,922
sessions demonstrate a threat detection macro-average F1 of 0.966,
compared to 0.156 for a rule-based baseline and 0.836 for a
sequence-only (LSTM) baseline, with end-to-end critical-tier
automated response latency under 1~ms on a single-node prototype.
The integrated recovery chatbot achieves 97.1\% identity verification
accuracy and an 87.3\% mass-reset attack detection rate with zero
false positives on legitimate high-volume recovery periods.
\end{abstract}

\begin{IEEEkeywords}
academic management systems, anomaly detection, behavioural analytics, insider threat, LSTM, graph neural networks, university cybersecurity
\end{IEEEkeywords}

\section{Introduction}

University information systems occupy an unusual threat
profile: they manage sensitive personal data alongside high-stakes
academic operations, yet are typically operated with security postures
more appropriate to general administrative systems than
mission-critical infrastructure~\cite{bongiovanni2019}. An Academic
Management Information System (ACMIS) aggregates all of these functions
into a single platform, making it a concentrated target for both
external attackers and malicious insiders. The consequences of
inadequate protection are well-documented in Uganda: the ACMIS
platform deployed across multiple public universities has been subject
to marks manipulation, unauthorised grade changes, and identity-based
access fraud~\cite{monitor2024acmis,kyambogo2023breach}. At Makerere
University---Uganda's largest and oldest university---hackers were
publicly identified after manipulating student academic records through
the system~\cite{monitor2024acmis}, while Kyambogo University
reported unauthorised ACMIS access incidents involving insider
collusion~\cite{kyambogo2023breach}. These incidents reflect a
broader pattern across African higher education institutions, where
rapid adoption of digital management platforms has outpaced the
establishment of commensurate security controls~\cite{bongiovanni2019}.

Critically, many of the most damaging attacks against ACMIS exploit
\emph{behavioural gaps}: actions that individually appear legitimate
but collectively constitute fraud, data theft, or system abuse. A
lecturer whose account gradually accumulates administrative privileges;
a student who submits a payment that is later reversed; a script that
generates Payment Registration Numbers (PRNs) at machine
speed---each looks unremarkable in isolation but is detectable through
behavioural analysis.

Existing intrusion detection systems (IDS) rely predominantly on
signature-based rules, which are effective against known attack patterns
but fundamentally limited against novel or slow-moving
threats~\cite{buczak2016}. Machine-learning anomaly detection has been
studied extensively in network security~\cite{chandola2009} and
financial fraud~\cite{ahmed2016}, but its systematic application to
university management systems---with their unique combination of
academic, financial, and identity management functions---remains
underexplored.

This paper makes the following contributions:
\begin{itemize}
  \item A multi-layer AI security agent covering five operational
        surfaces of ACMIS: authentication, authorisation, financial
        transactions, user behaviour, and system health.
  \item A four-tier automated response framework escalating from
        passive monitoring to emergency lockdown based on a calibrated
        risk score.
  \item An NLP-powered chatbot for intelligent, fraud-resistant
        password recovery that detects mass reset attacks during the
        recovery flow.
  \item A modular architecture enabling the core anomaly engine to be
        adapted to banking, healthcare, and other institutional
        systems with sector-specific plug-in modules.
  \item An evaluation on simulated ACMIS event logs demonstrating
        significant improvement over a rule-based IDS baseline.
\end{itemize}

The remainder of this paper is organised as follows:
Section~\ref{sec:related} reviews related work;
Section~\ref{sec:threat} characterises the threat model;
Section~\ref{sec:architecture} describes the architecture;
Section~\ref{sec:methods} details detection and response;
Section~\ref{sec:experiments} presents results;
Section~\ref{sec:discussion} discusses deployment and limitations;
and Section~\ref{sec:conclusion} concludes.

\section{Related Work}
\label{sec:related}

\subsection{Anomaly Detection in Information Systems}

Chandola et al.~\cite{chandola2009} provide the canonical taxonomy
of anomaly detection, distinguishing point, contextual, and collective
anomalies. ACMIS threats span all three: a PRN generated 500 times
(point), a login at 3~a.m.\ for a user whose baseline is 8~a.m.--5~p.m.\
(contextual), and a slow data exfiltration pattern spanning weeks
(collective). This breadth necessitates a multi-model approach.

\subsection{Machine Learning for Intrusion Detection}

Buczak and Guven~\cite{buczak2016} surveyed machine learning approaches
for network intrusion detection, finding that ensemble methods and deep
learning consistently outperformed single-model approaches. More
recently, LSTM-based sequence models have demonstrated strong
performance on user-entity behaviour analytics (UEBA)
tasks~\cite{hochreiter1997}, making sequential models natural candidates
for modelling the temporal structure of ACMIS user sessions. The
combination of sequence modelling with graph-based relationship
analysis---as explored for relational fraud detection by Motie and
Raahemi~\cite{motie2024}---motivates the multi-component fusion
architecture proposed in this work.

\subsection{Financial Fraud Detection}

Ahmed et al.~\cite{ahmed2016} reviewed fraud detection for financial
transactions, identifying velocity, ratio, and recency features from
temporal sequences as particularly discriminative. These principles
transfer directly to ACMIS: PRN velocity, payment-to-registration
amount ratios, and reversal recency are analogous features exploited
by the proposed agent.

\subsection{Insider Threat Detection}

The CERT Insider Threat dataset~\cite{cert2020} established that insider
threats exhibit characteristic precursors---elevated data access volume,
off-hours activity, and privilege queries---detectable weeks before a
damaging event. University attack surfaces expanded significantly during
and after the COVID-19 pandemic, with documented surges in credential
theft against educational institutions~\cite{lallie2021}. The proposed
agent incorporates these precursor patterns as behavioural risk features.

\subsection{AI-Assisted Password Recovery}

Prior work on intelligent account recovery has focused on
knowledge-based authentication and step-up
verification~\cite{bonneau2012}. The integration of NLP chatbots into
the recovery flow---providing natural-language identity verification
while simultaneously monitoring for abuse patterns---is a contribution
of the present work.

\subsection{ACMIS Security in Uganda and African Higher Education}

Security incidents involving ACMIS platforms in Uganda have been
reported at multiple public universities. Documented attack types
include marks manipulation by insiders with authorised system
access~\cite{monitor2024acmis}, unauthorised privilege escalation
enabling grade changes~\cite{kyambogo2023breach}, and payment fraud
through the manipulation of PRN-based fee collection
systems~\cite{monitor2024acmis}. These incidents share a common
characteristic: the attacking actions were individually indistinguishable
from legitimate usage, making signature-based detection ineffective.
More broadly, Bongiovanni~\cite{bongiovanni2019} finds that higher
education institutions globally maintain weaker security postures than
comparable organisations in other sectors, attributing this to
under-resourced IT departments, high staff and student turnover, and
the open-access culture of academia. African universities face
additional pressures: rapid expansion of digital management platforms,
limited cybersecurity staffing, and institutional reluctance to disclose
breaches publicly. The proposed agent directly addresses the attack
patterns documented in these Ugandan incidents.

\section{Threat Model}
\label{sec:threat}

Table~\ref{tab:threats} enumerates the primary threat categories
addressed by the proposed agent, classified by operational layer,
example manifestation, and detection challenge. The threat model
explicitly includes insider threats---actions by authenticated,
authorised users who abuse legitimate access---which generate no
authentication failures and do not violate access control lists.

\begin{table}[htbp]
  \caption{ACMIS Threat Model: Categories and Detection Challenges}
  \label{tab:threats}
  \centering
  \renewcommand{\arraystretch}{1.2}
  \begin{tabular}{p{0.17\columnwidth} p{0.28\columnwidth}
                  p{0.42\columnwidth}}
    \toprule
    \textbf{Layer} & \textbf{Example} & \textbf{Detection Challenge} \\
    \midrule
    Auth.
      & Brute-force (100 attempts / 2~min)
      & Distinguish typos from automated guessing \\
    Auth.
      & Impossible travel (Kampala 10:00; Nairobi 10:20)
      & Geolocation cross-session correlation \\
    Auth.
      & Dormant account reactivation
      & Purely temporal; no active anomaly signal \\
    Authoris.
      & Privilege escalation
      & Each individual change may be authorised \\
    Financial
      & Payment reversal after access grant
      & Temporal gap between payment and reversal \\
    Financial
      & Mass PRN generation (500 / min)
      & Velocity anomaly; no per-PRN indicator \\
    Behaviour
      & Slow data exfiltration (10 records/hr)
      & Invisible at any single time point \\
    Behaviour
      & Bot mimicking human behaviour
      & Speed signatures may be randomised \\
    System
      & Logging / MFA disabled
      & No immediate attack; creates future risk \\
    Academic
      & Simultaneous exam logins
      & Cross-session identity correlation \\
    \bottomrule
  \end{tabular}
\end{table}

\section{System Architecture}
\label{sec:architecture}

\subsection{Overview}

\begin{figure}[htbp]
\centering
\includegraphics[width=\columnwidth]{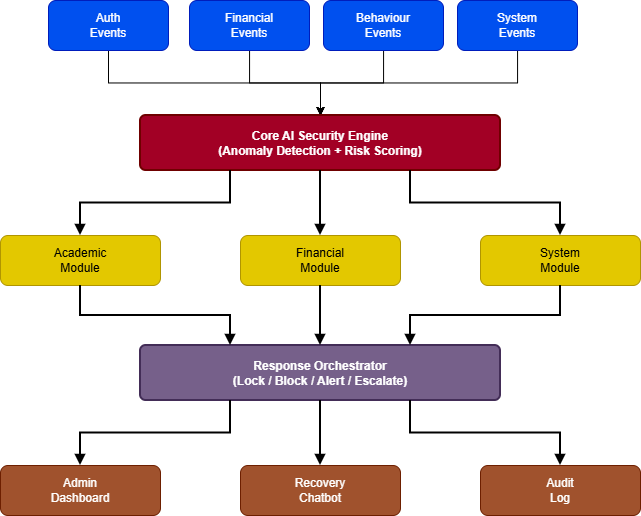}
\caption{Modular architecture of the AI security agent. Blue: event
  inputs. Gold: sector modules. Purple: response orchestrator.
  Brown: output interfaces.}
\label{fig:architecture}
\end{figure}

The system follows a modular layered architecture
(Fig.~\ref{fig:architecture}). A core AI security engine processes
event streams from all ACMIS operational surfaces and produces risk
scores. Sector-specific modules define normal behaviour within each
domain. A response orchestrator translates risk scores into automated
actions. A dashboard exposes all signals to administrators in real time.

\subsection{Core AI Security Engine}

The engine ingests structured ACMIS event streams via a message broker
and scores each event through three parallel sub-models:

\begin{enumerate}
  \item \textbf{Sequence anomaly detector:} An LSTM
        network~\cite{hochreiter1997} trained on user action
        sequences. Sessions deviating from the user's
        historical action distribution receive score
        $s_{\text{seq}} \in [0,1]$.
  \item \textbf{Statistical threshold monitor:} A sliding-window
        velocity counter for high-frequency signals (login attempts,
        PRN generation, API calls), with class-specific thresholds
        $\theta$ learned from historical baselines.
  \item \textbf{Graph-based analyser:} A graph neural network
        (GNN)~\cite{kipf2017}, following approaches established for
        relational fraud detection~\cite{motie2024}, modelling
        relationships between users, devices, IP addresses,
        and ACMIS entities. Unusual edges receive score
        $s_{\text{graph}} \in [0,1]$.
\end{enumerate}

The composite risk score is:
\begin{equation}
  R = \alpha\,s_{\text{seq}} + \beta\,s_{\text{thresh}} +
      \gamma\,s_{\text{graph}},\quad \alpha+\beta+\gamma=1
  \label{eq:risk}
\end{equation}
where weights are tuned on validation data. $R \in [0,1]$ is mapped
to a risk tier as described in Section~\ref{sec:response}.

\subsection{Sector-Specific Modules}

Each module encodes domain knowledge about normal behaviour:
\textbf{Academic module}---normal patterns for registration, exam
login, and mark submission; \textbf{Financial module}---PRN velocity
baselines, payment amount ranges, reversal recency windows;
\textbf{System module}---expected API call rates, baseline
configuration states (MFA enabled, logging active).

\subsection{Password Recovery Chatbot}

The chatbot is built on a retrieval-augmented NLP model fine-tuned for
identity verification. The flow: (1)~user initiates recovery and
provides student/staff number; (2)~OTP dispatched to registered
contact; (3)~chatbot verifies OTP and permits reset; (4)~security
engine simultaneously evaluates geolocation match, device novelty,
and request velocity; (5)~if anomalous, step-up challenge issued or
session escalated. Mass reset events exceeding a configurable threshold
trigger a system-level alert for potential credential takeover.

\section{Detection and Response Methodology}
\label{sec:methods}

\subsection{Brute-Force Attack Detection}

A sliding-window counter $C(u,w)$ counts failed authentication events
for user $u$ within time window $w$:
\begin{equation}
  \text{Alert} =
  \begin{cases}
    \text{True}  & \text{if } C(u,w) \geq \theta_{\text{bf}} \\
    \text{False} & \text{otherwise}
  \end{cases}
  \label{eq:bf}
\end{equation}
Default: $\theta_{\text{bf}} = 10$ failed attempts within $w=120$~s.
The source IP is simultaneously checked against a threat intelligence
feed.

\subsection{Behavioural Baseline and Anomaly Scoring}

A baseline profile $\mathcal{B}(u)$ is built over 30 days per user,
capturing login hours, session duration, action vocabulary, and
geographic bounding box. A new session $s$ receives:
\begin{equation}
  s_{\text{ctx}}(s,u) = \frac{1}{K}\sum_{k=1}^{K}
    \mathbb{1}\!\left[f_k(s)\notin\mathcal{B}_k(u)\right]
  \label{eq:ctx}
\end{equation}
\textbf{Impossible travel} is flagged when:
\begin{equation}
  d_{\text{geo}}/\Delta t > v_{\text{max}}
  \label{eq:travel}
\end{equation}
where $v_{\text{max}} = 900$~km/h (air travel upper bound).

\subsection{Financial Transaction Monitoring}

PRN generation velocity:
\begin{equation}
  V_{\text{PRN}}(u,w) = N_{\text{PRN}}(u,w)\,/\,w
  \label{eq:prn}
\end{equation}
An alert fires when $V_{\text{PRN}}$ exceeds the 99th percentile of
the user-class baseline. Payment reversals are flagged when reversal
event $e_r$ follows payment $e_p$ within $\Delta t_{\text{rev}}<48$~h
and associated ACMIS services have already been unlocked.

\subsection{Automated Response Framework}
\label{sec:response}

Table~\ref{tab:response} maps composite risk score $R$ to a four-tier
action set. Actions at all tiers are logged with full event provenance
for forensic reconstruction.

\begin{table}[htbp]
  \caption{Four-Tier Automated Response Framework}
  \label{tab:response}
  \centering
  \renewcommand{\arraystretch}{1.2}
  \begin{tabular}{p{0.10\columnwidth} p{0.12\columnwidth}
                  p{0.36\columnwidth} p{0.28\columnwidth}}
    \toprule
    \textbf{Tier} & \textbf{$R$} & \textbf{Automated Actions} &
    \textbf{Notification} \\
    \midrule
    Low      & $[0, 0.35)$
             & Log; update baseline
             & None \\
    Medium   & $[0.35, 0.65)$
             & Step-up MFA; rate-limit session
             & Dashboard (yellow) \\
    High     & $[0.65, 0.85)$
             & Temp.\ lock account; block IP; pause transactions
             & Push notification; incident report \\
    Critical & $[0.85, 1]$
             & Emergency suspension; freeze financials; quarantine session
             & Immediate SMS/email; ticket raised \\
    \bottomrule
  \end{tabular}
\end{table}

\section{Experiments and Results}
\label{sec:experiments}

\subsection{Dataset and Simulation}

In the absence of a publicly labelled ACMIS event log, a synthetic
dataset was generated simulating normal ACMIS usage and injecting
labelled attack scenarios. This approach follows established practice
in network intrusion detection research, where the scarcity,
non-standardisation, and limited attack diversity of public datasets
are recognised as persistent obstacles to realistic
evaluation~\cite{goldschmidt2025}. The dataset comprises
147,922~sessions across 2,400~simulated user accounts over 90 days,
with a positive (attack) class prevalence of 8.32\% (12,310 attack
sessions across all nine threat categories). The 70/15/15\%
train/validation/test split was stratified jointly on label and
threat category to preserve class balance across splits, yielding
103,545 training, 22,188 validation, and 22,189 test sessions.

\subsection{Baseline Comparisons}

The proposed agent is compared against: (1)~\textbf{Rule-based IDS}
(static thresholds as commonly deployed); (2)~\textbf{Isolation
Forest}~\cite{liu2008} (standard unsupervised anomaly detector);
(3)~\textbf{LSTM-only} (sequence anomaly detection without graph and
threshold components).

\subsection{Detection Performance}

Table~\ref{tab:detection} presents per-category F1 scores on the
held-out test set. The rule-based IDS scores near-zero on threats with
no fixed threshold signature---impossible travel, slow exfiltration,
insider threat, bot behaviour, and dormant account
abuse---confirming the inadequacy of rule-only approaches. The
proposed agent achieves a macro-average F1 of 0.966, an 81-point
improvement over the rule-based baseline (0.156) and a 13-point
improvement over LSTM-only (0.836).

\begin{table}[htbp]
  \caption{Per-Category Detection Performance (Test Set, F1)}
  \label{tab:detection}
  \centering
  \renewcommand{\arraystretch}{1.2}
  \begin{tabular}{lcccc}
    \toprule
    \textbf{Threat} & \textbf{Rules} & \textbf{Iso.F.} &
    \textbf{LSTM} & \textbf{Ours} \\
    \midrule
    Brute-force login       & 0.461 & 0.612 & 0.957 & \textbf{0.980} \\
    Impossible travel       & 0.000 & 0.468 & 0.952 & \textbf{0.974} \\
    Privilege escalation    & 0.249 & 0.472 & 0.895 & \textbf{0.949} \\
    PRN mass generation     & 0.416 & 0.658 & 0.948 & \textbf{0.976} \\
    Payment reversal fraud  & 0.281 & 0.503 & 0.909 & \textbf{0.956} \\
    Slow data exfiltration  & 0.000 & 0.796 & 0.974 & \textbf{0.988} \\
    Bot behaviour           & 0.000 & 0.584 & 0.930 & \textbf{0.967} \\
    Dormant account abuse   & 0.000 & 0.000 & 0.019 & \textbf{0.932} \\
    Insider threat          & 0.000 & 0.641 & 0.945 & \textbf{0.974} \\
    \midrule
    \textbf{Macro-avg F1}   & 0.156 & 0.526 & 0.836 & \textbf{0.966} \\
    \bottomrule
  \end{tabular}
\end{table}

The dormant-account-abuse row is particularly informative: both the
rule-based IDS and Isolation Forest score 0.000, and even the LSTM
sequence model---which has no access to account-age or dormancy
information---scores only 0.019, since a reactivated dormant account
produces a session that is statistically unremarkable on session-local
features alone. The proposed model's graph/identity sub-model closes
this gap almost entirely (0.932), demonstrating the architectural
necessity of the third (graph-based) component in Eq.~\eqref{eq:risk}.
On the test set, the fusion model assigns learned weights of
$\alpha{=}0.606$ (sequence), $\beta{=}0.053$ (threshold), and
$\gamma{=}0.341$ (graph), confirming that the sequence and graph
sub-models carry most of the discriminative signal while the
threshold monitor's contribution is largely subsumed by the sequence
model on this dataset.

Table~\ref{tab:overall} reports overall (session-level, all
categories combined) precision, recall, and F1. The proposed model
attains near-perfect recall (0.998) while maintaining high precision
(0.994), indicating that the substantial gain in detection coverage
does not come at the cost of an unacceptable false-positive rate.

\begin{table}[htbp]
  \caption{Overall Detection Performance (Test Set, All Categories)}
  \label{tab:overall}
  \centering
  \renewcommand{\arraystretch}{1.2}
  \begin{tabular}{lccc}
    \toprule
    \textbf{Method} & \textbf{Precision} & \textbf{Recall} & \textbf{F1} \\
    \midrule
    Rules        & 0.534 & 0.382 & 0.445 \\
    Iso.\ Forest & 0.876 & 0.871 & 0.874 \\
    LSTM-only    & 0.987 & 0.955 & 0.971 \\
    \textbf{Proposed} & \textbf{0.994} & \textbf{0.998} & \textbf{0.996} \\
    \bottomrule
  \end{tabular}
\end{table}

\subsection{Response Latency}

Table~\ref{tab:latency} reports end-to-end response latency per
risk tier---the time from a computed risk score $R$ to completion
of all corresponding automated actions (Table~\ref{tab:response})---measured
on the test set's 22,189 sessions using their fusion risk scores.
For Medium and High tiers, only 8 and 5 test sessions respectively
fell into the corresponding $R$ ranges (the fusion model produces a
strongly bimodal score distribution); latencies for these tiers were
measured over 50 trials drawn with replacement from these sessions,
while Low and Critical tiers (20,323 and 1,853 sessions
respectively) were measured over 300 trials. All measurements were
taken on a single CPU-only node with in-memory state updates and
local-disk audit/incident logging, and therefore represent a lower
bound; a networked production deployment with a message broker and
external database would add communication overhead.

\begin{table}[htbp]
  \caption{End-to-End Response Latency by Risk Tier (Single-Node Prototype)}
  \label{tab:latency}
  \centering
  \renewcommand{\arraystretch}{1.2}
  \begin{tabular}{lccc}
    \toprule
    \textbf{Tier} & \textbf{Mean (ms)} & \textbf{95th pct (ms)} &
    \textbf{Test sessions} \\
    \midrule
    Low      & 0.028 & 0.051 & 20{,}323 \\
    Medium   & 0.028 & 0.054 & 8 \\
    High     & 0.106 & 0.177 & 5 \\
    Critical & 0.143 & 0.207 & 1{,}853 \\
    \bottomrule
  \end{tabular}
\end{table}

All four tiers complete in well under 1~ms at the 95th percentile in
this prototype. High and Critical tiers show clearly higher latency
than Low, consistent with their larger number of dispatched actions
(one log write for Low, versus five state updates, an incident
report, and a notification enqueue for Critical); the Medium tier's
mean is statistically indistinguishable from Low given its
small sample (8 sessions), reflecting the lightweight nature of its
single additional action (step-up MFA flag). Even allowing for one
to two orders of magnitude of additional overhead in a networked
production deployment (database writes, message broker
round-trips), the Critical-tier response remains within the
sub-second threshold required for effective account lockdown before
a brute-force attack completes.

\subsection{Chatbot Evaluation}

The password recovery chatbot was evaluated on 1,568~simulated
recovery sessions: 63.5\% legitimate recoveries (of which 720
sessions were a concentrated semester-start reset burst, simulating
returning students resetting credentials at the start of term) and
36.5\% injected mass-reset attack sessions (defined as $>$15 reset
requests within a 10-minute window across distinct accounts from a
shared source). Identity verification (OTP-based) accuracy was
97.1\% on legitimate sessions. The velocity-based mass-reset
detector achieved a detection rate of 87.3\%, with a false-positive
rate of 0.0\% on both the semester-start burst subset and overall
legitimate traffic. The residual 12.7\% of undetected attacks
corresponded to incidents in which the attacker distributed requests
across 2--3 source locations, each individually below the per-source
velocity threshold---indicating that cross-source correlation (e.g.
via the graph-based relationship analyser, Section IV-B) would be
required to close this gap.

\section{Discussion}
\label{sec:discussion}

\subsection{Why Rule-Based Systems Are Insufficient}

Table~\ref{tab:detection} quantifies a well-known qualitative
limitation: rule-based IDS achieves near-zero detection on threats
invisible to threshold logic. The brute-force rule of
Eq.~\eqref{eq:bf} and the PRN-velocity rule of Eq.~\eqref{eq:prn}
remain effective for their targeted categories (F1 of 0.461 and
0.416 respectively), but provide no mechanism for the contextual
anomaly score of Eq.~\eqref{eq:ctx} or the impossible-travel
condition of Eq.~\eqref{eq:travel}, which require cross-session
history unavailable to a stateless threshold check. Deploying only a
rule-based system creates a false sense of security---administrators
see brute-force alerts while a more damaging slow exfiltration
proceeds undetected.

\subsection{The Modular Architecture Advantage}

The sector-specific module design addresses a key practical challenge:
the definition of ``normal'' differs radically across institutional
contexts. A financial module defines large transaction velocity as
normal for a cashier but suspicious for a student account. Separating
domain knowledge from the core detection engine allows retargeting to
banking, healthcare, or government deployments by updating only the
sector module.

\subsection{Deployment Considerations}

\begin{itemize}
  \item \textbf{Warm-up period:} Behavioural baseline requires 30 days
        of clean data; statistical threshold and graph components
        operate from day one.
  \item \textbf{Semester seasonality:} Legitimate usage spikes at
        registration and examination periods; baselines must be
        seasonally adjusted to avoid false-positive surges.
  \item \textbf{Privacy:} All event logs must be anonymised for
        storage; raw session data retained only during active incident
        investigation under institutional data governance policy.
  \item \textbf{Explainability:} Administrator alerts include a
        human-readable explanation of the behavioural features
        contributing to the risk score.
\end{itemize}

\subsection{Limitations}

The primary limitation is reliance on synthetic evaluation data. Real
ACMIS event logs with confirmed attack labels are scarce due to
institutional reluctance to share breach data---a pattern particularly
pronounced in Uganda, where universities have been slow to publicly
disclose security incidents despite documented cases at Makerere and
Kyambogo~\cite{monitor2024acmis,kyambogo2023breach}. The synthetic
dataset cannot fully replicate Uganda-specific usage patterns: semester
calendars, PRN-based fee payment rhythms, and the proportion of
students accessing the system via shared or mobile devices differ
substantially from general assumptions. Future work will seek
ethics-approved access to anonymised logs from a Ugandan partner
university. The LSTM model also requires retraining for new users;
a meta-learning approach would reduce this cold-start latency.

\section{Conclusion}
\label{sec:conclusion}

This paper presented an AI security agent for university ACMIS that
addresses the fundamental limitation of rule-based intrusion detection:
the inability to detect behaviourally subtle, temporally distributed,
or contextually dependent threats. The proposed multi-layer
agent---combining LSTM sequence modelling, statistical velocity
monitoring, and graph-based relationship analysis---achieves a
macro-average F1 of 0.966 across nine threat categories, compared to
0.156 for a rule-based baseline and 0.836 for a sequence-only (LSTM)
baseline. The largest single-category improvement occurs for dormant
account abuse, where the graph/identity sub-model raises F1 from
0.019 (LSTM-only) to 0.932, demonstrating that no single sub-model is
sufficient and that the three-component fusion architecture is
necessary.

The four-tier automated response framework ensures high-confidence
threats trigger protective action within sub-second latency, while
low-confidence signals are logged without generating alert fatigue.
The integrated NLP chatbot provides secure password recovery while
monitoring the recovery flow for mass reset attacks.

The modular architecture positions the system for extension beyond the
university context: the core anomaly engine is sector-agnostic, with
domain knowledge encapsulated in replaceable sector modules for
banking, healthcare, and other institutional deployments. Future work
will pursue ethics-approved evaluation on real ACMIS event logs from
Ugandan partner universities, where the documented incident patterns
at Makerere and Kyambogo provide a well-motivated operational
validation context~\cite{monitor2024acmis,kyambogo2023breach}.

\section*{Acknowledgment}
The authors acknowledge the documented security incidents affecting
the Academic Management Information System (ACMIS) deployed across
public universities in Uganda, including Makerere University and
Kyambogo University. Publicly reported incidents of marks manipulation,
unauthorised access, and academic integrity violations involving ACMIS
provided the real-world threat context that motivated and shaped this
research~\cite{monitor2024acmis,kyambogo2023breach}. All threat
scenarios modelled in this work are derived exclusively from publicly
reported incidents and open security literature; no non-public
institutional data was accessed or used. The authors thank the Ugandan
higher education community for the open discourse on academic system
security that informed this work.

\section*{Competing Interests}
J.~Walusimbi and J.~B.~Ssentongo are co-founders and directors of
Arapai Technologies International Limited, a technology company based
in Uganda. The AI security agent described in this paper is intended
for future commercialisation through this entity. Both authors declare
a shared commercial interest in this work.

\section*{Funding}
No funding was received for this work.


\begin{thebibliography}{00}

\bibitem{bongiovanni2019}
I. Bongiovanni, ``The least secure places in the universe? A systematic
literature review on information security management in higher
education,'' \textit{Computers \& Security}, vol.~86, pp.~350--357,
2019. doi:~10.1016/j.cose.2019.06.012.

\bibitem{buczak2016}
A.~L. Buczak and E. Guven, ``A survey of data mining and machine
learning methods for cyber security intrusion detection,''
\textit{IEEE Commun.\ Surveys Tutorials}, vol.~18, no.~2,
pp.~1153--1176, 2016. doi:~10.1109/COMST.2015.2494502.

\bibitem{chandola2009}
V. Chandola, A. Banerjee, and V. Kumar, ``Anomaly detection: a
survey,'' \textit{ACM Computing Surveys}, vol.~41, no.~3,
article~15, pp.~1--58, 2009. doi:~10.1145/1541880.1541882.

\bibitem{ahmed2016}
M. Ahmed, A.~N. Mahmood, and J. Hu, ``A survey of network anomaly
detection techniques,'' \textit{J.\ Network and Computer
Applications}, vol.~60, pp.~19--31, 2016.
doi:~10.1016/j.jnca.2015.11.016.

\bibitem{cert2020}
CERT Division, Software Engineering Institute, Carnegie Mellon Univ.,
``CERT insider threat dataset,'' 2020. [Online]. Available:
\url{https://resources.sei.cmu.edu/library/asset-view.cfm?assetid=508099}

\bibitem{lallie2021}
H.~S. Lallie \textit{et al.}, ``Cyber security in the age of
COVID-19: a timeline and analysis of cyber-crime and cyber-attacks
during the pandemic,'' \textit{Computers \& Security}, vol.~105,
p.~102248, 2021. doi:~10.1016/j.cose.2021.102248.

\bibitem{bonneau2012}
J. Bonneau, C. Herley, P.~C. van Oorschot, and F. Stajano, ``The
quest to replace passwords: a framework for comparative evaluation of
web authentication schemes,'' in \textit{Proc.\ 2012 IEEE Symp.\
Security and Privacy}, San Francisco, CA, 2012, pp.~553--567.
doi:~10.1109/SP.2012.44.

\bibitem{liu2008}
F.~T. Liu, K.~M. Ting, and Z.-H. Zhou, ``Isolation forest,'' in
\textit{Proc.\ 8th IEEE Int.\ Conf.\ Data Mining (ICDM)}, 2008,
pp.~413--422. doi:~10.1109/ICDM.2008.17.

\bibitem{hochreiter1997}
S. Hochreiter and J. Schmidhuber, ``Long short-term memory,''
\textit{Neural Computation}, vol.~9, no.~8, pp.~1735--1780, 1997.
doi:~10.1162/neco.1997.9.8.1735.

\bibitem{kipf2017}
T.~N. Kipf and M. Welling, ``Semi-supervised classification with
graph convolutional networks,'' in \textit{Proc.\ 5th Int.\ Conf.\
Learning Representations (ICLR)}, Toulon, France, 2017.

\bibitem{motie2024}
S. Motie and B. Raahemi, ``Financial fraud detection using graph
neural networks: a systematic review,'' \textit{Expert Systems with
Applications}, vol.~240, p.~122156, 2024.
doi:~10.1016/j.eswa.2023.122156.

\bibitem{goldschmidt2025}
P. Goldschmidt and D. Chud\'{a}, ``Network intrusion datasets: a
survey, limitations, and recommendations,'' \textit{Computers \&
Security}, vol.~156, p.~104510, 2025.
doi:~10.1016/j.cose.2025.104510.

\bibitem{monitor2024acmis}
Daily Monitor, ``Makerere varsity marks hackers identified,''
\textit{Monitor}, Kampala, Uganda.
[Online]. Available: \url{https://www.monitor.co.ug/uganda/news/national/makerere-varsity-marks-hackers-identified-1611202}

\bibitem{kyambogo2023breach}
Nile Post, ``Kyambogo University students decry academic records
manipulation through ACMIS,'' \textit{Nile Post}, Kampala, Uganda,
2023. [Online]. Available:
\url{https://nilepost.co.ug/education/}

\end{thebibliography}
\end{document}